\documentclass{article}
\usepackage{ijcai21}

\usepackage{times}
\usepackage{soul}
\usepackage{url}
\usepackage[hidelinks]{hyperref}
\usepackage[utf8]{inputenc}
\usepackage[small]{caption}
\usepackage{graphicx}
\usepackage{amsmath}
\usepackage{amsthm}
\usepackage{booktabs}
\usepackage{algorithm}
\usepackage{algorithmic}
\usepackage{subfigure}
\usepackage{multirow}
\usepackage{todonotes}
\usepackage{longtable}
\usepackage{tabularx}
\usepackage{tikz}
 
\newcommand*\halfcirc[1][1ex]{%
  \begin{tikzpicture}
  \draw[fill] (0,0)-- (90:#1) arc (90:270:#1) -- cycle ;
  \draw (0,0) circle (#1);
  \end{tikzpicture}}
\newcommand*\fullcirc[1][1ex]{\tikz\fill (0,0) circle (#1);} 

\newcolumntype{b}{>{\hsize=.95\hsize}X}
\newcolumntype{s}{>{\hsize=.35\hsize}X}
\newcolumntype{m}{>{\hsize=.6\hsize}X}

\newcommand{\paraspace}{\vspace{0.05in}}
\newcommand{\parab}[1]{\paraspace\noindent{\bf #1} }

\newenvironment{icompact}{
  \begin{list}{$\bullet$}{
    \parsep 1pt plus 1pt
    \partopsep 1pt plus 1pt
    \topsep 1pt plus 2pt minus 1pt
    \itemsep 1.5pt plus 1pt
    \parskip 0pt plus 2pt
    \leftmargin 0.15in}
       }
  {\normalsize\end{list}}

\title{Aegis: A Trusted, Automatic and Accurate Verification Framework for Vertical Federated Learning}

\author{
Cengguang Zhang$^1$
\and
Junxue Zhang$^{1,2}$
\and
Di Chai$^{1,2}$
\and
Kai Chen$^{1,3}$
    \affiliations
    $^1$iSING Lab, Hong Kong University of Science and Technology\\
    $^2$Clustar Technology Co., Ltd.\\
    $^3$Peng Cheng Lab
    \emails
    \{czhangch, jzhangcs, dchai, kaichen\}@cse.ust.hk
}

\begin{document}

\maketitle

\begin{abstract}
Vertical federated learning (VFL) leverages various privacy-preserving algorithms, e.g., homomorphic encryption or secret sharing based SecureBoost, to ensure data privacy. However, these algorithms all require a semi-honest secure definition, which raises concerns in real-world applications. In this paper, we present Aegis, a trusted, automatic, and accurate verification framework to verify the security of VFL jobs. Aegis is separated from local parties to ensure the security of the framework. Furthermore, it automatically adapts to evolving VFL algorithms by defining the VFL job as a finite state machine to uniformly verify different algorithms and reproduce the entire job to provide more accurate verification. We implement and evaluate Aegis with different threat models on financial and medical datasets. Evaluation results show that: 1) Aegis can detect $95\%$ threat models, and 2) it provides fine-grained verification results within $84\%$ of the total VFL job time.

\end{abstract}

\section{Introduction}
 
Vertical federated learning (VFL), or feature-based federated learning, has recently attracted increasing research and industrial interest since its first paper proposed by WeBank in 2019~\cite{yang2019federated}. Unlike Google's horizontal federated learning that targets co-training a model by securely aggregating parameters from a local model trained over homogeneous datasets on massive mobile devices, VFL co-trains a model across heterogeneous datasets that share the same data ID. These datasets are usually owned by several companies or institutions. For example, company A has user online behavior data $X_1$; company B has user credit, consumption data $X_2$ and label $Y$; they can combine the two datasets and increase the conversion rate of financial products purchases by using VFL. To support the computation paradigm of VFL, WeBank open-sources the most widely adopted VFL framework: FATE. Our paper mainly focuses on the FATE but should be applicable to other VFL frameworks.
 

VFL enables security data cooperation among different data silos by leveraging several secure protocols, e.g., a privacy-preserving vertical tree-based model through homomorphic encryption and secret sharing in~\cite{wu2020privacy}. These protocols guarantee that no one can reveal the data of other parties. However, existing privacy-preserving protocols typically require the semi-honest secure definition~\cite{yang2019federated}, indicating that the system is privacy-preserving only when the participants all follow the pre-defined protocol. However, the semi-honest assumption is hard to be guaranteed in reality and malicious participants exist. These malicious parties will not follow the secure protocol, making the VFL system vulnerable to a broader set of attacks, e.g., the malicious participants could use model poisoning to dirty the jointly trained model, which brings significant performance reduction.~(\S\ref{sec:motivation threat model})

Therefore, we propose a new notion of VFL verification. However, VFL verification mainly faces three challenges:
1) The VFL party is easy to be compromised or imitated by malicious insiders. Thus it is challenging to guarantee the credibility of the verification framework if deployed at the end.
2) It is difficult for the verification framework to adapt to the continuously evolving privacy-preserving algorithms in VFL.
3) As various random algorithms are used in VFL to further protect privacy, ensuring the framework's verification accuracy under such randomness is challenging.
Our framework supports two modes, real-time verification and postponed verification, to balance the trade-off between efficiency and security. Real-time verification is low-latency but cannot detect some of the attacks, and postponed verification takes more time but can give a comprehensive ensure on security.
 
In this paper, to address the above challenges, we build a trusted, automatic and accurate verification framework: Aegis~(\S\ref{sec:design choices}). Aegis has the following design choices:
 \begin{icompact}
 	\item To prevent the malicious attackers on the endhost, our framework is integrated with the L-7 gateway, which can not be accessed by VFL parties. We collect data from the gateway and store it in a non-tamperable storage system such as blockchain, ensuring the security of collected data.
 	\item To adapt to the rapidly evolving VFL algorithm, we use a finite state machine to describe a VFL algorithm generally. Then we design a module to automatically generate finite state machine and data rule tables for different algorithms. Besides, we reproduce the VFL process to verify the sent data from the local party.
	\item To reproduce the original job process, we store job metadata, including training data, model configuration files, code, homomorphic encryption key pairs, model parameters and random seeds. Therefore, we can reproduce the job accurately and reduce the overhead, which includes communication and calculation time.
\end{icompact}

From the perspective of system architecture, Aegis is composed of three modules: Aegis Core, Message Collector, and VFL Analyzer. The Aegis Core module is used as an intermediate scheduling module to coordinate the work of Message Collector and VFL Analyzer and manage their life cycle~(\S\ref{sec:design Aegis Core}). Message Collector collects the transmission data and the data is verified by VFL Analyzer~(\S\ref{sec:design Message Collector}, \S\ref{sec:design VFL Analyzer}).
 
We implement Aegis based on FATE. We use Kubernetes to manage the modules and use the spring MVC framework to provide external RESTful APIs. We take FATE as an example, integrate Message Collectors in Exchange. Messages are collected and preprocessed in Message Collector and stored according to downstream application requirements in different storage systems. Aegis uses a finite state machine for automatic verification. For data verification, we use rule tables and local replication experiments to verify the security~(\S\ref{sec:implementation}).
 
We built some representative threat models for security analysis and used our framework for verification. The evaluation results show that our framework detects most of the potential security risks but can not detect the attacks, such as model poisoning~(\S\ref{sec:eval security}). In addition, we conducted sufficient experiments on different datasets to describe the overhead while using Aegis in real-time verification. We also demonstrate the overhead reduction of reproducing VFL jobs locally for different models~(\S\ref{sec:eval overhead}).
\vspace{-2ex}
\section{Background \& Motivation}
\subsection{Vertical Federated Learning Process}
\label{sec:motivation VFLP}
We first introduce the concept and workflow of VFL based on FATE, which is an industry-level VFL platform. Vertical federated learning applies when two datasets share the same sample ID space but differ in feature space. VFL is the process of aggregating these different features and computing the training losses and gradients in a privacy-preserving manner to build a model with data from both parties collaboratively\cite{yang2019federated}.
The recent concentration point of VFL research is overcoming statistical challenges and promote security. Although the VFL algorithm theoretically guarantees data privacy security, in the actual industrial landing, there are still many system-level security risks that need to be considered. For example, pre-trained models from other organizations may provide a backdoor for adversary~\cite{wang2020backdoor}, and insiders of parties may collude with the other malicious parties to leak data or poison the model. The increment of VFL algorithms and frameworks exacerbated these problems. To better analyze these problems, we start with the analysis of the VFL process. 

\begin{figure}[H]
\centering
\includegraphics[width=0.5\textwidth]{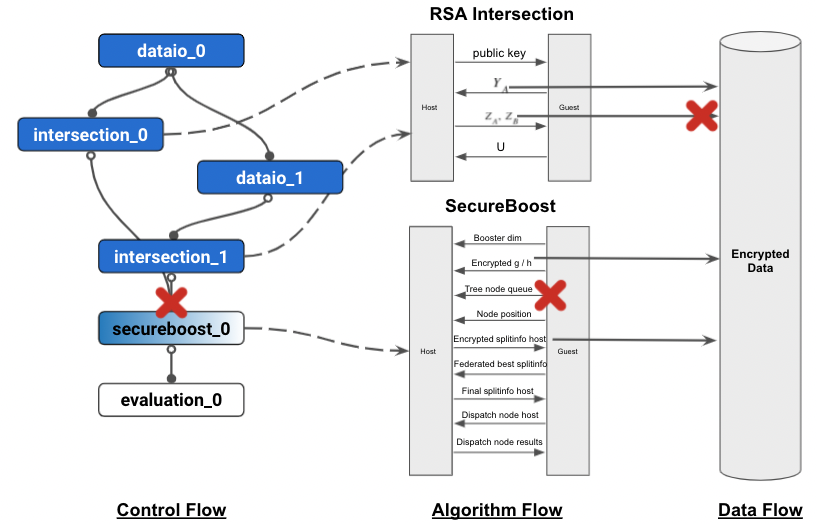}
\caption{Vertical Federated Learning Process}
\label{Fig.1}
\vspace{-2ex}
\end{figure}

As shown in Figure 1, we define the whole VFL process with three different granularities of message flow, control flow, algorithm flow, and data flow. When we start a VFL job, the job can be described as a directed acyclic graph(DAG). Between each component of the DAG, there are control flow messages to deliver the control command of the process, such as submit a job or report progress. Each component of the DAG is a subtask of VFL job, such as intersection or SecureBoost. Inside the component, there are algorithm flow messages to synchronize variables. The most fine-grained message flow is the data flow inside each algorithm flow message, which transfers the encrypted data between parties.
 \vspace{-1.5ex}
\subsection{Threat Model}
\label{sec:motivation threat model}
Based on the above analysis, we have categorized VFL's exchanging messages into three levels: control flow, algorithm flow, and data flow. Following, we organize the existing attack models according to our categorization and briefly introduce how to perform an attack under each message level.
\vspace{-1.5ex}
\begin{icompact}
	\item \parab{Control Flow Attack} The malicious party can send specific control messages in the control flow to attack VFL jobs. For example, the malicious adversary can prohibit the model from converging to the best performance by sending an early "stop" control message.
	\vspace{-1.5ex}
	\item \parab{Algorithm Flow Attack} The algorithm flow highly depends on the machine learning models. The exchanging messages in algorithm flow are carefully designed not to leak private information (e.g., protected by homomorphic encryption) and keep a minimum amount to reduce the communication burden. The overall logic is called the secure protocol. The malicious parties will not follow the secure protocol and modify the transmitted message to perform an attack. For example, people in the guest party can be corrupted by the malicious parties and send extra messages that contain the samples' label, which is the most private data at the guest party.
	\vspace{-1.5ex}
	\item \parab{Data Flow Attack} Apart from violating the pre-defined protocol, the malicious participants can also attack the system through model poisoning and data poisoning. Existing studies have shown that model poisoning is more powerful than data poisoning \cite{cao2020fltrust}, and most data poisoning attacks will eventually take effect by dirtying the model parameters. Thus we only consider model poisoning in the data flow attack. The malicious parties can use the poisoning attacks to perform a targeted \cite{bagdasaryan2020backdoor,xie2019dba} (e.g., misleading the model to predict green cars into birds) or untargeted attack \cite{fang2020local} (e.g., simply poison the model such that performance in each label is reduced). 
\end{icompact}

It is worth noting that the above attacks happen when there are malicious participants in the system, and all the works that proved to be secure under semi-honest assumptions will face the above attack challenges. Thus a verification framework to detect malicious activities and enhance the system's privacy-preservation in the real-world application is the need of the day.

\subsection{Challenge}
\label{sec:motivation challenge}
Our goal is to design a trusted, automatic and accurate framework for VFL job verification. Thus we mainly face the following challenges:
\begin{icompact}
 	\item When there are malicious attackers on the end host, how to ensure that the framework is trustworthy and difficult to hack.
 	\item With the rapid development of federated learning, a variety of new algorithms emerge in an endless stream, and a variety of algorithms are also integrated in FATE, such as FedAvg\cite{mcmahan2017communication}, FedVision\cite{liu2020fedvision}, SecureBoost\cite{DBLP:journals/corr/abs-1901-08755} and so on. How to ensure that the framework automatically adapts to most algorithms in this case.
 	\item Various random algorithms are used in federated learning to ensure randomness. How to overcome randomness in postponed verification to ensure that the original experiment is reproduced?
 \end{icompact}
\vspace{-2ex}
\section{Aegis}
\subsection{Overview of Aegis}
\label{sec:design choices}
\paragraph {Design rationality} 
Our design is based on the following observations. First, most of the VFL frameworks deployed in industrial applications are based on a star topology. A star topology is a topology in which all nodes are individually connected to a central connection node. Therefore all messages will be forwarded through the central point, the gateway, or the router. We assume that the gateway is secure and can not be accessed by VFL parties. From this observation, Aegis is designed to integrate with the gateway. We collect messages from the gateway and store them in a non-tamperable storage system such as blockchain, which ensures the security of collected messages. In this way, we not only make Aegis trusted but also convenient to collect messages. 

Secondly, although VFL algorithms vary from each other, the whole process can be represented by a finite state machine(FSM). Thus we use an FSM to decouple the VFL algorithm and verification. Furthermore, we design a module to automatically generate finite state machine and data rule tables for different algorithms in verification, which make Aegis can automatically adapt to different VFL algorithms.

Last but not least, the randomness in the VFL algorithm is provided by random variables, such as random seeds and random masks, which means if we fix the random variable, the VFL job can be perfectly reproduced. For data flow verification, we store random seeds for data sampling, shuffle, encryption, and decryption, to reproduce the VFL job and provide accurate verification.

The whole framework contains three main components: \textbf{Aegis Core}, \textbf{Message Collector}, \textbf{VFL Analyzer}. The Aegis core model mainly acts as a coordinator of the whole framework. The Message Collector is integrated with the gateway, which collects, preprocesses, and classifies messages. The VFL Analyzer verifies the security of VFL jobs.
\vspace{-2ex}
\begin{figure}[H]
\centering
\includegraphics[width=0.5\textwidth]{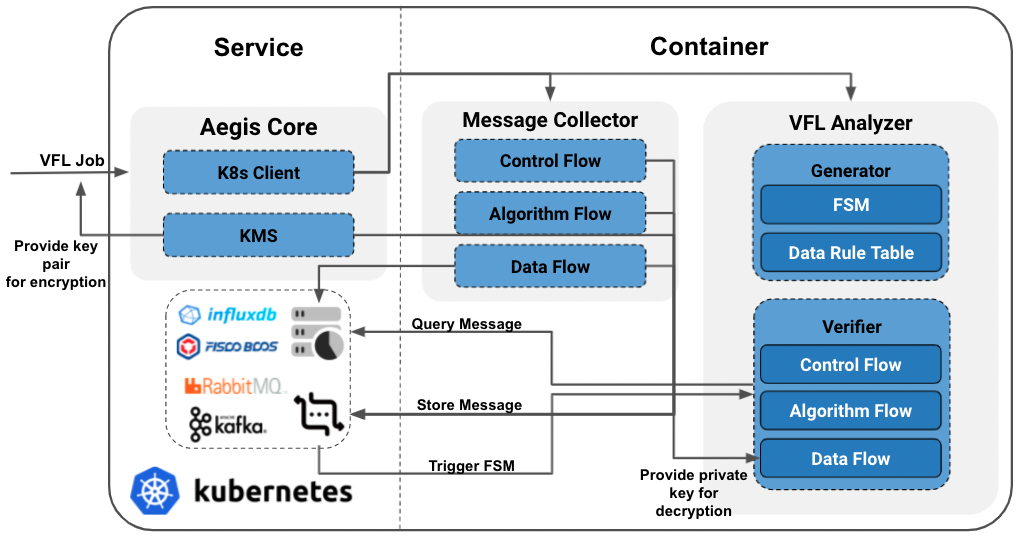}
\vspace{-3ex}
\caption{Aegis Workflow}
\label{Fig.2}
\vspace{-3ex}
\end{figure}
\paragraph{Workflow}
Figure 2 demonstrates Aegis's workflow to verify a VFL job. After the federated learning job is initiated, the process will initiate a request to the KMS service in the Aegis Core module and obtain a secret key pair to encrypt the data. Then, the communication traffic of the federated learning job is forwarded through the L-7 gateway. Thus, the Message Collector in the gateway will collect the transmission data, classify and preprocess the data according to the data transmission method, and store it in different storage according to the needs of downstream applications In the system. Furthermore, in VFL Analyzer, on the one hand, the process is verified according to the FSM generated in advance by the federated learning job. On the other hand, the private key is obtained from the KMS service for decryption, and the data is verified according to the rule table. In the postponed verification, we will also reproduce the training process based on the collected training-related data, such as code, model parameters, training data, random seeds, etc., and compare the transmitted data to ensure security.
 \vspace{-1.5ex}
\subsection{Aegis Core}
\label{sec:design Aegis Core}
First of all, the Aegis Core module will create and manage the gateway through Kubernetes and supervise the life cycle of the gateway. Furthermore, it provides a KMS for each VFL task to generate and manage the secret key used in the task, facilitating us to decrypt the data in the VFL Analyzer for verification. When a VFL task is started, the Message Collector on the gateway will report the new task to the Aegis Core module. Then the Aegis Core module will create the corresponding VFL Analyzer according to the user-customized verification level. The results of the VFL Analyzer verification will be reflected in real-time. If it is a postponed verification, users can also start the VFL Analyzer through the Aegis Core to achieve it.
\begin{figure}[H]
\centering
\includegraphics[width=0.36\textwidth]{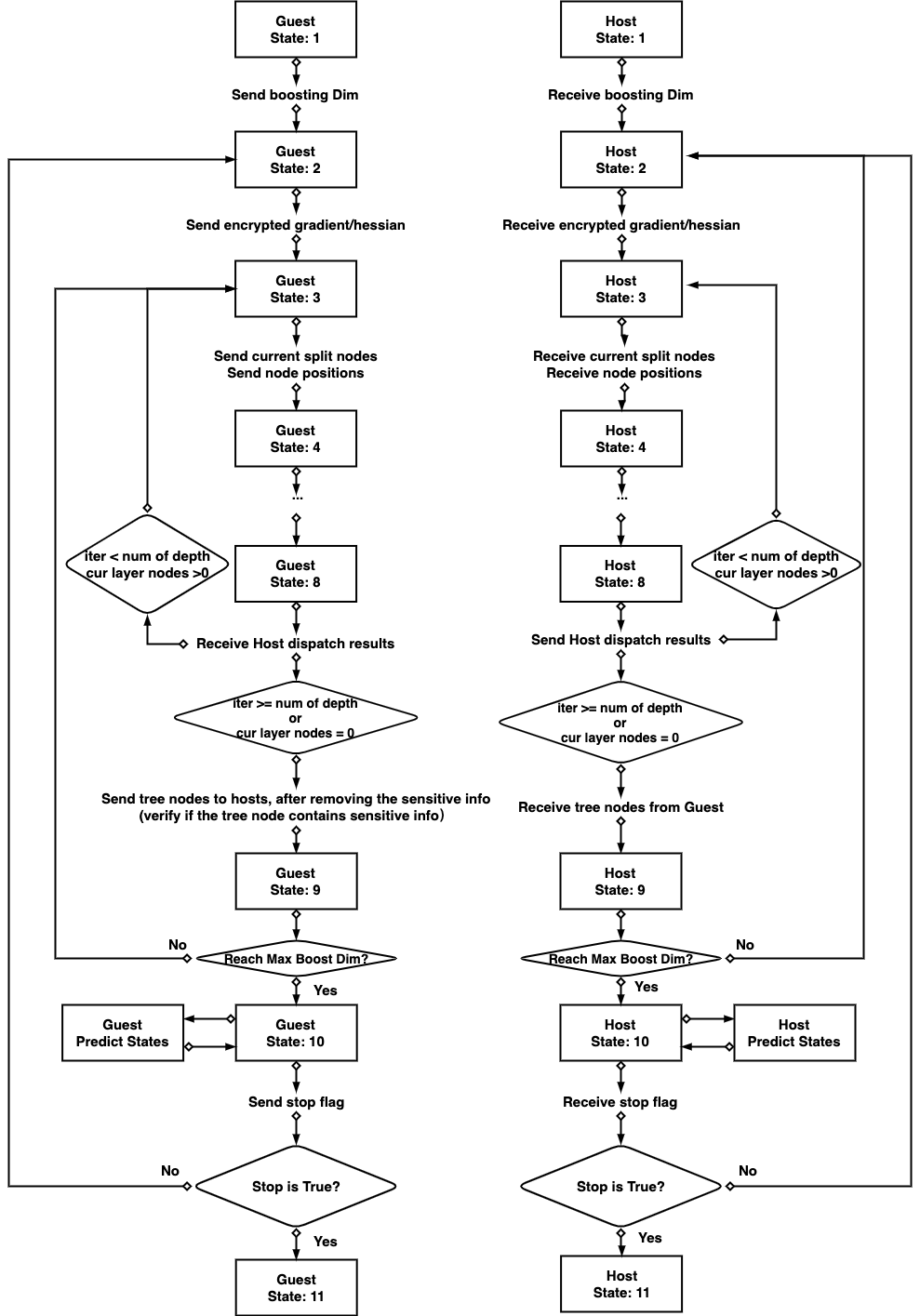}
\caption{FSM Example: Secure Boost}
\label{Fig.2}
\vspace{-2ex}
\end{figure}
\parab{Key Management Service} 
The secret key plays a vital role in cryptographic encryption and affects the security of the encryption algorithm. Therefore, the secret keys need to be properly managed. Most VFL frameworks will generate secret keys during VFL jobs and discard the secret keys after use. However, we need to obtain the corresponding secret keys after running the encryption algorithm to decrypt and verify the data. In this case, KMS is added to our system. VFL jobs generate and obtain secret keys through the KMS API. In addition, KMS can also provide functions such as updating the secret keys regularly, setting the expiration date of the secret keys, etc.
 \vspace{-1.5ex}
\subsection{Message Collector}
\label{sec:design Message Collector}
 Message Collector is integrated with a gateway, especially in the FATE framework with an exchange, and collects the transferred data before forwarding. But these messages need to be preprocessed and classified by the Message Collector. For example, in the FATE framework, we can divide the message based on the transfer method. The control flow messages are transferred by gRPC unary call, while the algorithm flow and data flow messages are transferred by gRPC stream call. After that, some of the messages will be divided into several partitions, so Message Collector should reassemble the messages and filter out the useless metadata. The messages classified as control flow, algorithm flow, and data flow are stored in different databases according to the requirements for efficiency and security. If real-time verification is required, then we will store the message in the message queue. If it is postponed verification, we store the data in a time-series database or blockchain storage.
 \vspace{-1.5ex}
\subsection{VFL Analyzer}
\label{sec:design VFL Analyzer}
In this section, we will illustrate the key part of verification, named VFL Analyzer, which is formed by two parts, \textbf{Generator} and \textbf{Verifier}. 

\parab{Generator}
The generator mainly includes two parts: a finite state machine generator for control flow and algorithm flow verification, and the other is a rule table generator for data flow verification. The operation of a VFL job is carried out according to a DAG. We automatically generate the different states of the \textbf{FSM} through each node in the DAG and automatically generate the state machine transactions according to the dependencies between each node of the DAG. For the generation of the finite state machine of the algorithm flow, the user can generate it by providing the config file or by running the algorithm once in a safe environment with a small data set. After that user can obtain a template config file and make minor modifications such as the number of loop variables to generate the config file. Data may vary depending on the algorithm. Therefore, we will generate \textbf{Data Rule Table} to check the attributes of encrypted data. We can obtain some data attributes by static analyzing the code, such as data size, data type, etc. Besides, the rule table can be customized by users to check extra attributes of data.

\parab{Verifier}
Verifier is divided into three parts, respectively, to verify control flow, algorithm flow, and data flow. Control flow between different algorithms is not very different, so we set a unified verification process. On the other hand, the algorithm and data flow vary greatly with different algorithms, so we use the results obtained in the Generator for verification. To verify control flow and algorithm flow, we abstract the entire process as a finite state machine, a message or a combination of multiple messages as an event, which can trigger a state transition. Each time the verification of the algorithm flow starts from an initial state, the state of the state machine changes with the message flow. When an abnormal state occurs, the algorithm flow verification will alarm and trigger the pre-set handling methods. If the state machine runs as expected, the VFL process can pass the verification of the algorithm flow. Take algorithm flow as an example, Secure Boost algorithm can be abstracted like the state machine diagram shown in Figure 3. We can obtain the states, events, and transactions based on the diagram.

To verify data flow, we will first request a private key corresponding to the VFL job to decrypt the data. Then, encrypted data attributes will be checked according to the rule table. Finally, we will reproduce the VFL job locally, refer to the training data and metadata, and compare the results with the transmitted data to ensure the security of the VFL job.

\vspace{-2ex}
\section{Implementation}
\label{sec:implementation}
In the specific implementation of this system, we use Kubernetes to manage our various components, which is also suitable for large-scale distributed federated learning scenarios. The system follows the micro-service architecture, making our framework low-coupling, extensible, and easy to maintain. We expose RESTful API for external access, such as FATE and UI, and use gRPC API for internal access.
 \vspace{-1.5ex}
\subsection{Aegis Core}
We implement the Aegis Core module based on the Spring MVC framework, mainly including key management service(KMS) and Aegis Core service.
\paragraph{KMS} 
We provide user registration, login, authorization, and authentication services. We will provide the generated JSON Web Token (JWT) for subsequent authentication when the user logs in. Two types of authentications are applied with different granularities. The coarse-grained authentication is verifying if the user is registered, and the fine-grained authentication uses the Access Control List(ACL) to authenticate the user permission of each key. The creator of the secret key will grant user permissions. The verification department is a superuser and has access to all secret keys.
\paragraph{Aegis Core service}
We implemented authorization, authentication, and basic CURD operations for Message Collectors and VFL Analyzers. In addition, the Aegis Core service is connected to the Kubernetes API server to manage the underlying Message Collector and VFL Analyzer containers. Users can easily manage the Message Collectors and VFL Analyzers with RESTful API in the Aegis Core service. The lifecycle of Message Collectors and VFL Analyzers will be monitored and updated with gRPC API by Aegis Core service.
 \vspace{-1.5ex}
\subsection{Message Collector} 
This module is integrated with the gateway, specifically, FATE exchange. In FATE exchange, we make minor modifications to the code both in gRPC unary call and stream call, add different Message Collectors and call the collect method before each forwarding to collect messages. In the FATE framework, the data is divided into multiple partitions according to the number of processors, which perform calculations during the training process. Therefore, after classifying the data according to the metadata and transfer method, we also need to reassemble the transferred data to facilitate subsequent algorithm flow and data flow verification. Finally, we store the preprocessed data in different databases for different usage in chronological order. In particular, we use RabbitMQ and Kafka as our message queue in real-time verification. While for postponed verification, we use influxDB and FISCO BCOS to ensure data security and immutability. We will store the hash value in blockchain for verification and other original data in influxDB for VFL reproduction.
 \vspace{-1.5ex}
\subsection{VFL Analyzer}
\paragraph{Generator}
We provide users with multiple ways to generate FSM. One is to write configuration files following our prescribed format directly. The second is that the program automatically outputs a configuration file that conforms to the format, and the user can customize it based on the configuration file we generated. Automatic generation is divided into static and dynamic generation. The static method is to add annotations when transmitting data and entering the loop. We can output configuration files that conform to the format through static analysis of the code. The dynamic method is to run a small-scale experiment in a trusted environment, obtain the configuration file based on the data transmitted, and then obtain the correct number of loops based on the loop variable in the code. For the generation of the rule table, we also obtain the characteristics of the transmission variable through static code analysis and generate the corresponding rule table. Users can customize more check items based on our rule table.

\paragraph{Verifier}
Regarding verification, we will divide it into two parts: real-time and postponed. In the real-time part, we will first use the finite state machine to verify the correctness of the control flow and algorithm flow, whether there is any private information, and destroy the training information. Secondly, we will verify the correctness of some meta-information of the encrypted data, such as data length, data type, etc. In the postponed part, we will run the VFL job with a trusted code, the random variable of the job will be obtained from a database. In the training process, we record the sent messages and get remote data collected in immutable storage. Finally, after model reproduction, we will verify the hash value of the model and sent messages. 
\vspace{-2ex}
\begin{table*}[t]
\small
\centering
\begin{tabularx}{\textwidth}{llllc}
\toprule
Message Flow & Attack & Adversaries & Verification Mechanisms & Results \\
\midrule
\multirow{3}*{Control} & A01: FATE Client Impersonation & Insider & Control Flow FSM Verification & \fullcirc[0.5ex] \\
~ & A02: Control Flow Tampering & Insider, Outsider & Control Flow FSM Verification & \fullcirc[0.5ex] \\
~ & A03: Private Data Leakage & Insider & Control Flow FSM Verification  & \fullcirc[0.5ex] \\
\hline
\multirow{2}*{Algorithm} & A03: Private Data Leakage & Insider & Algorithm Flow FSM Verification  & \fullcirc[0.5ex] \\
~ & A04: Algorithm Flow Tampering & Insider, Outsider & Algorithm Flow FSM Verification & \fullcirc[0.5ex] \\
\hline
\multirow{2}*{Data} & A03: Private Data Leakage & Insider & VFL job Repo & \fullcirc[0.5ex] \\
~ & A05: Code/Data Tampering & Insider & Data Rule Table + VFL job Repo & \fullcirc[0.5ex] \\
~ & A06: Model Poisoning & Insider, Outsider & Data Rule Table + VFL job Repo & \halfcirc[0.5ex] \\
~ & A07: Direct Remote Control & Insider, Outsider & Data Rule table & \fullcirc[0.5ex] \\
\hline
Other & A08: Attacking VFL Keys & Insider & KMS & \fullcirc[0.5ex] \\
\bottomrule
\end{tabularx}
\label{tab:table1}
\caption{Assessment of attacks on VFL job and respective countermeasures, following the attacker models defined in Section 5.}
\vspace{-3ex}
\end{table*}

\section{Evaluation}
\subsection{Security Analysis}

\label{sec:eval security}
We now discuss how adversaries could attempt to attack the VFL job. Table 1 summarizes the attacks and respective defense mechanisms. Finally, we discuss why the protection from the above adversaries implies the fulfillment of the security goal and therefore solves the initial challenges. Most attacks can be prevented successfully, while the attack, e.g., model poisoning, is hard to prevent unless we conduct a federated verification with other parties.

\parab{Adversary Types} Attacks can be carried out by insiders and outsiders. Insider attacks include those launched by the VFL server and the participants in the VFL system. Outsider attacks include those launched by the eavesdroppers on the communication channel and by users of the final VFL model when it is deployed as a service. Insider attacks are generally more dangerous than outsider attacks, as it strictly enhances the adversary's capability.\cite{lyu2020privacy}

\parab{A01: FATE Client Impersonation} Malicious processes in the local system pretend to be a FATE Client and use the channel between parties to exchange private messages. In the control flow FSM, our framework will verify the control command and therefore prevents such attacks.

\parab{A02: Control Flow Tampering} Malicious parties or outsiders can tamper with the control flow message, e.g., send a stop command to damage the training. This kind of attack can be prevented by the control flow FSM verification as well.

\parab{A03: Private Data Leakage} Private data can be revealed in each message flow. For example, malicious parties can send extra algorithm flow messages with private information to other parties, and the VFL job runs as expected. However, such privacy data leakage will be detected by our verification framework.

\parab{A04: Algorithm Flow Tampering} Algorithm flow messages can be distorted by the adversary, e.g., the stop flag in the VFL algorithm indicates the time to early stop VFL job. Thus, if the stop flag is tampered with by malicious parties, the performance of VFL will be affected. Our framework will check the algorithm flow messages using FSM verification, therefore prevents these attacks.

\parab{A05: Code/Data Tampering} Code and VFL data in the local system can be changed by the adversary, which significantly affects the VFL job. Therefore we use Data Rule Table to check the variable size and other features. Besides, we reproduce the VFL job with trusted code and data to prevent tampering.

\parab{A06: Model Poisoning} The model poisoning attacks could be performed by malicious outsiders or insiders corrupted by the adversary. Briefly, our system can reveal the poisoning activities during the joint training, e.g., modifying the model updates. Because our system logs all the intermediate results, and the participants could jointly perform a reproduction trial to verify the intermediate results. If any difference is found, then there are suspicious activities during the training. There are cases in which the attacks cannot be discovered. For example, the malicious adversary could also use a poisoned data at the beginning and pretends to be honest during the training. Thus our framework is partially secure regarding the poisoning attacks.

\parab{A07: Direct Remote Control} VFL platforms such as FATE have some system-level backdoors. For instance, the Pickle package in Python transfers the Python object to a byte array and sends the byte array to the receiver. If we convert some shell commands into a binary stream, the other party's system will execute the corresponding commands, which means the system could be controlled by the adversary remotely. Our framework decrypts and deserializes the data for verification, thereby defending against such attacks.

\parab{A08: Attacking VFL Keys} KMS provides a security guarantee for the entire life cycle of the key. Thus, users do not need to worry about key leakage during the entire cycle of key creation, update, and invalidation.

\subsection{Overhead Analysis}
We conducted sufficient experiments with Secure Boost and Logistic Regression models as examples which are the two most popular VFL models in real-world applications, to show that the delay of real-time verification is low, and the time of postponed verification is also within the acceptable range. \label{sec:eval overhead}

\parab{Experiment Settings} We use different datasets to test the overhead of our framework. Datasets details are shown in Table 2. We use Intel(R) Xeon(R) Silver 4114 CPU @ 2.20GHz and Ubuntu 18.04 for experiments. The average delay is about 50ms, and the average bandwidth is about 1Gbps, which roughly simulates a wide area network.
\vspace{-2ex}
\begin{table}[h!]
  \tiny
  \begin{center}
    \label{tab:table2}
    \begin{tabular}{c|c|c|c|l} 
    \toprule
      \textbf{Name} & \textbf{Guest Dim} & \textbf{Host Dim} & \textbf{Size} & \textbf{Description}
      \\
      \hline
      Small & 10 & 20 & 569 & Breast cancer diagnostic data\\
      \hline
      Medium & 20 & 80 & 5000 & Mock data\\
      \hline
      Large & 13 & 10 & 30000 & Credit card clients data\\
      \bottomrule
    \end{tabular}
  \end{center}
  \vspace{-3ex}
  \caption{Datasets Description}
\end{table}
\vspace{-3ex}

\parab{Real-time Verification Delay}
The delay of real-time verification is shown in Table 3. We can find from the data that the time to verify the control flow is generally shorter than that of the algorithm flow because the FSM of the control flow is simple, and the amount of messages for the control flow is also less. In addition, the real-time verification time is extremely short compared to the time of the overall task. For actual jobs, security problems can be found by Aegis within a very short delay. Noted that we use the small dataset because real-time verification overhead has little correlation with the dataset. However, even in the smallest dataset, the total VFL time greatly exceeds the real-time verification time, so our conclusion is still valid under a larger dataset.
\vspace{-2ex}
\begin{table}[h!]
  \scriptsize
  \begin{center}
    \label{tab:table3}
    \begin{tabular}{c|c|c|c} 
    \toprule
      \textbf{Algorithm} & \textbf{Control Flow} & \textbf{Algorithm Flow} & \textbf{FL job Time}\\
      \hline
      Secure Boost & 461ms & 593ms & 129s \\
      \hline
      Logistic Regression & 448ms & 648ms & 111s\\
      \hline
      RSA Intersection & - & 458ms & 6.47s\\
      \hline
      Raw Intersection & - & 432ms & 5.88s\\
      \bottomrule
    \end{tabular}
    \vspace{-2ex}
    \caption{Real-time Verification Overhead Analysis}
    \vspace{-4ex}
  \end{center}
\end{table}
\vspace{-2ex}

\parab{Reproduce Overhead Reduction}
Compared with the original task, the time to reproduce the job saves two parts of time. One part is the time consumed by communication, and the other part is the time waiting for other parties to calculate. We take two-party federated learning as an example. We, as a guest, analyze the changes in these two parts of time under different data sizes and the ratio of the host to guest data dimensions, $R_{h/g}$. It shows that in our framework when the amount of data and $R_{h/g}$ is larger, the amount of time saved will be longer, so the overall reproduction time will be within an acceptable range. Noted that our experiments are based on medium and large datasets, specifically, using different dimensions of host data in the medium dataset and different data sizes in large datasets.
\begin{figure}[H]
\vspace{-2ex}
\centering
\includegraphics[width=0.4\textwidth]{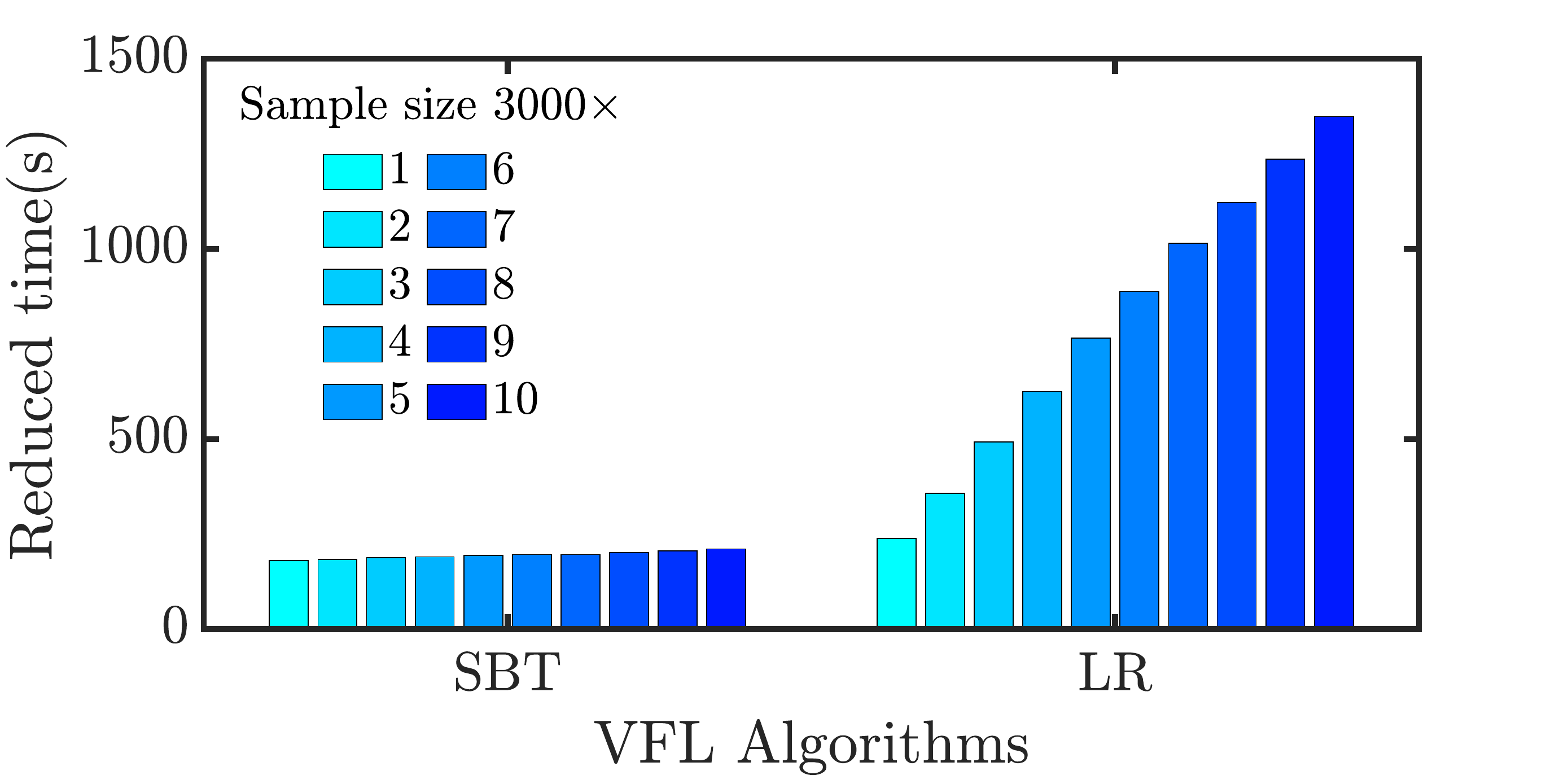}
\vspace{-2.5ex}
\caption{Communication overhead reduction for different dataset size}
\label{Fig.4}
\vspace{-4ex}
\end{figure}

\begin{figure}[H]
\vspace{-2ex}
\centering
\includegraphics[width=0.4\textwidth]{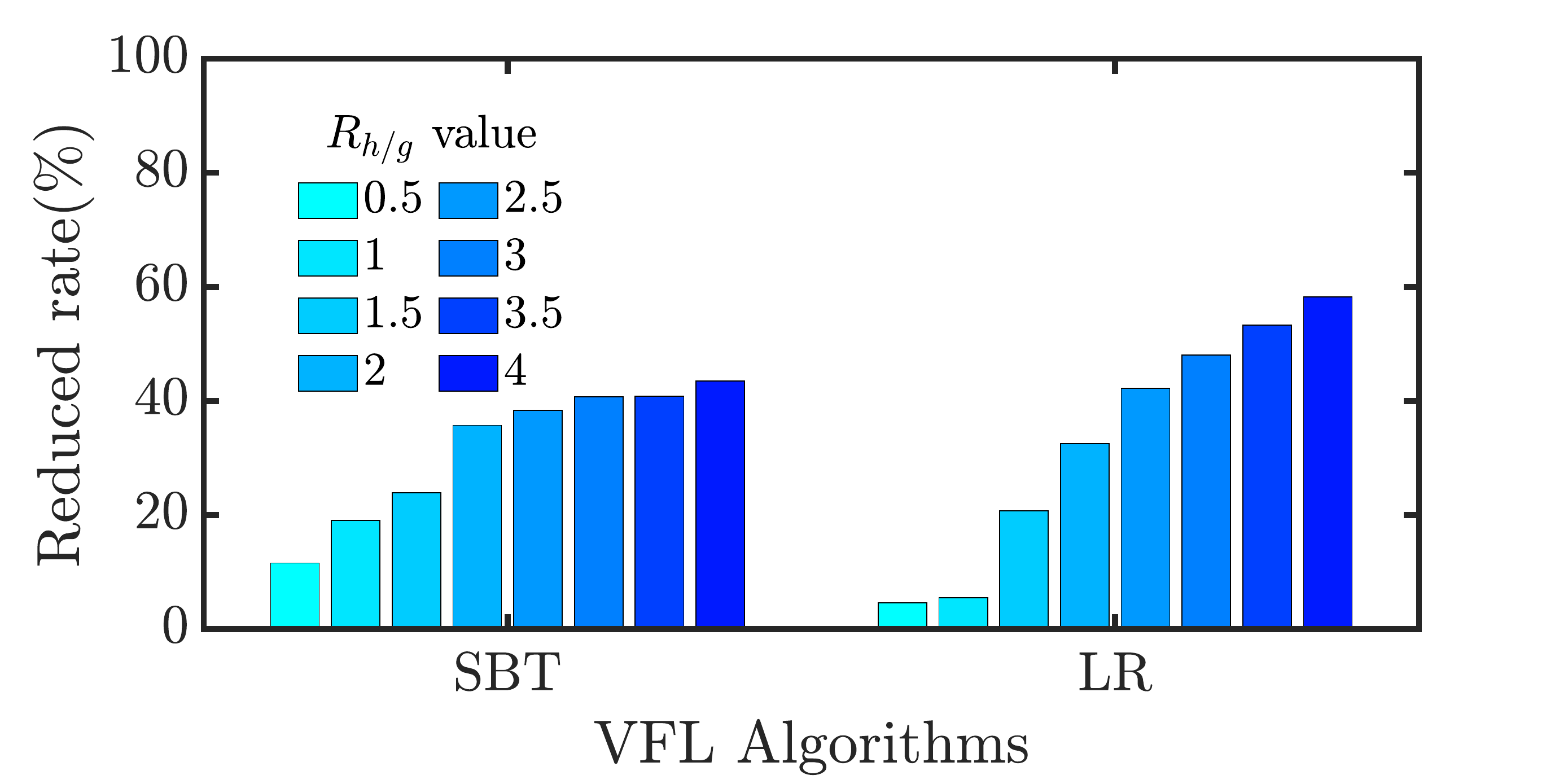}
\vspace{-2.5ex}
\caption{Calculation overhead reduction rate for different $R_{h/g}$}
\label{Fig.5}
  \vspace{-2ex}
\end{figure}
\section{Related Work}
\parab{Federated Learning} Federated learning aims to collaboratively train machine learning models across data silos without privacy leakage. Horizontal federated learning targets at securely training a model across homogenous samples on different mobile devices~\cite{chai2020secure,chai2021federated}. Later, vertical federated learning paradigm~\cite{yang2019federated}, targeting at training machine models across heterogenous data in different data silos is proposed.Currently, different VFL algorithms have been proposed, such as heterogenous logistic regression~\cite{yang2019federated}, SecureBoost~\cite{DBLP:journals/corr/abs-1901-08755}, FedRec~\cite{yang2020federated}, etc.

\parab{Poisoning Attacks and Defense Methods} Federated learning is vulnerable to poisoning attacks, in which the malicious parties could poison the local data or model during the joint training. \cite{bagdasaryan2020backdoor} and \cite{fang2020local} proposed model poisoning attacks, in which the local malicious participants modify the local updates before uploading or directly use dirty label in training. Meanwhile, methods against the poisoning attacks have also been proposed. \cite{cao2020fltrust} proposed a defense method in which the server maintains a trusted gradients and verifies the clients' uploaded updates by measuring the cosine similarity between the trusted gradients and the received updates.
\vspace{-2ex}
\section{Conclusion}
In this paper, we proposed Aegis, a trusted, automatic, and accurate VFL verification framework. Aegis can give a comprehensive verification within an acceptable time. This work will promote the practical use of the VFL algorithm, thereby promoting the development of federation learning.
\vspace{-2ex}

\bibliographystyle{named}
\bibliography{ijcai21}

\end{document}